\documentclass[%
 reprint,
showpacs,%
 amsmath,amssymb,
 aps,prl,
floatfix
]{revtex4-1}
\usepackage[usenames]{color}
\usepackage{subfigure}
\usepackage{graphicx} %
\usepackage{todonotes}

\newcommand{\mc}[1]{{\color{black}#1}}
\begin{document}

\title{Anharmonic and Quantum Fluctuations
in Molecular Crystals: \\
A First-Principles Study of the Stability of Paracetamol}

\author{Mariana Rossi}
\thanks{These two authors contributed equally to the work.}
\author{Piero Gasparotto\normalfont\textsuperscript{*}}

\author{Michele Ceriotti}
\email{michele.ceriotti@epfl.ch}
\affiliation{Laboratory of Computational Science and Modeling, Institute of Materials, {\'E}cole Polytechnique F{\'e}d{\'e}rale de Lausanne, 1015 Lausanne, Switzerland}%

\date{\today}

\begin{abstract}
Molecular crystals often exist in multiple 
competing polymorphs, showing significantly different
physico-chemical properties. 
Computational crystal structure prediction is
key to interpret and guide the search for 
the most stable or useful form: A real challenge 
due to the combinatorial search space, and the 
complex interplay of subtle
effects that work together to determine the relative
stability of different structures. 
Here we take a comprehensive approach based on
different flavors of thermodynamic integration in order
to estimate all contributions to the free energies of 
these systems with density-functional theory, including 
the oft-neglected anharmonic contributions and 
nuclear quantum effects.
We take the two main stable forms of paracetamol as a 
paradigmatic example.
We find that anharmonic contributions,
different descriptions of van der Waals interactions,
and nuclear quantum effects all matter to
 quantitatively determine
the stability of different phases. 
Our analysis highlights the many challenges inherent in the
development of a quantitative and predictive framework
to model molecular crystals.
However, it also indicates which of the components
of the free energy can benefit from a cancellation of
errors that can redeem the predictive power of approximate models, 
and suggests simple steps that could be taken to 
improve the reliability of \emph{ab initio} crystal structure prediction.
\end{abstract}


\footnotetext{\dag~Electronic Supplementary Information (ESI) available: [details of any supplementary information available should be included here]. See DOI: 10.1039/b000000x/}

\maketitle 

Many compounds of biological and pharmaceutical
importance crystallize as molecular crystals.
Thus, predicting the stability of different polymorphs 
is of paramount importance, as the crystal structure 
can affect the  solubility and pharmacokinetics of 
drugs \cite{cheney,Kipp2004109}.
In fact, there have been several incidents with drugs 
having to be recalled \cite{RotigotineRecall2008},
production plants having to be halted \cite{Ritonavir},
or patents being disputed \cite{Cefadroxil}
for reasons connected to the drug's polymorphism.
In this field, the \mc{synergy between 
experiments and  computational crystal structure
prediction (CSP)} promises to eliminate these
inconveniences. However, the combinatorial complexity
of the configuration space\cite{day+09acb}, as well as the need of 
evaluating relative stabilities of polymorphs with exquisite accuracy due to their very similar binding 
free energies, pose formidable obstacles to 
achieving this goal. In fact, this challenge has been recognized by 
the community: The Cambridge Crystallographic Data Centre 
has launched a series of blind tests 
focused on molecular crystal predictions, in order to assess the predictive
power of various modelling approaches \cite{Bardwell:2011, Groom:bm5086}.
Here, we will show
\mc{for the paradigmatic example
of paracetamol}
that although an accurate potential energy is 
necessary for the reliable
modelling of these structures, 
anharmonic (quantum) free energy contributions
are just as important. \mc{The development of accurate methods
to evaluate these terms will be one of the major
challenges to enable application of CSP to 
larger and more flexible compounds.}

\begin{figure}[htbp]
\centering
\includegraphics[width=0.35\textwidth]{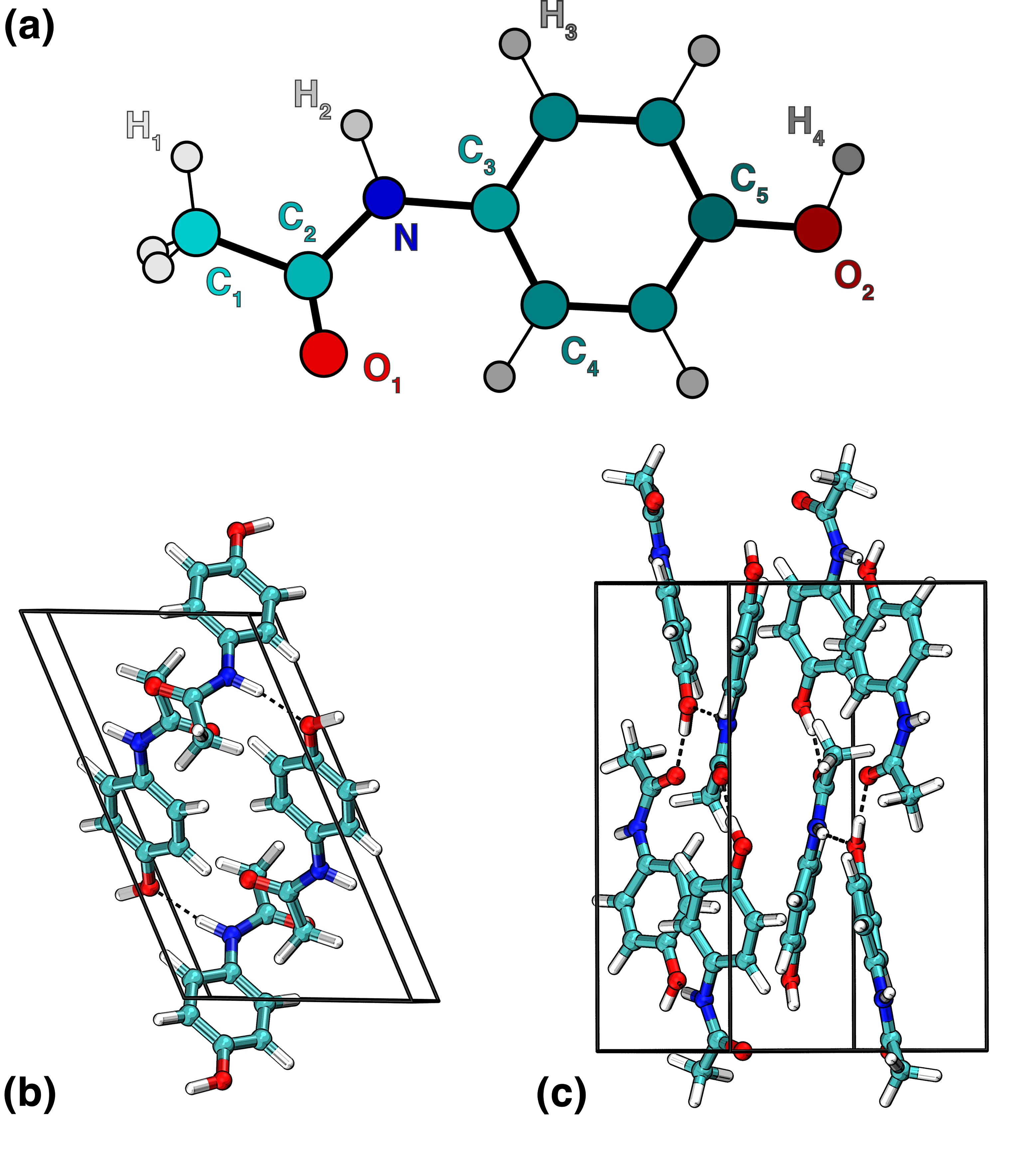}
\caption{(a) The paracetamol molecule including the color code that we use throughout the paper for each atom. Carbons are in shades of green, oxygens in shades of red, nitrogen is blue, and hydrogens are in shades of grey. The crystal structures of paracetamol in (b) monoclinic form I with four molecules per unit cell and (c) orthorhombic form II with eight molecules per unit cell. }
\label{fig:para}
\end{figure}

In this work we take a comprehensive approach involving
several different types of thermodynamic integration in order
to evaluate the full \textit{ab initio}
binding free energies of molecular crystals, disentangling
the many different contributions -- harmonic, anharmonic,
classical and quantum mechanical -- that contribute to the
free energies of these structures.
We choose as an example
Paracetamol (N-acetyl-p-aminophenol), a molecule of particular
interest because of its wide use as an antipyretic and analgesic. 
We consider the monoclinic form I (fI) and the othorhombic form II (fII)  \cite{HaisaMaeda1974, HaisaMaeda1976} (shown in Fig.\ref{fig:para}), which are the most common for this compound.
They are predicted to be similarly energetically stable (differences are estimated to be between 10 and 60 meV/molecule), with form I being favored at room temperature and pressure \cite{Espeau2005524, BeyerPrice:2001hs, NeumannPerrin2009,DistasioTkatchenko2012}. 

It is known that the relative stability of different forms can depend
on the delicate interplay of qualitatively different physical effects, as has
been recently demonstrated in the case of Aspirin \cite{MaromTkatchenko2013, ReillyTkatchenko2014},
where focus was given to van der Waals (vdW) forces and harmonic entropic terms.
Other aspects that have received less attention so far
are the contribution to free-energies
that comes from nuclear quantum effects (NQEs), as well as the evaluation
of temperature-related anharmonic effects, that have 
been proven necessary to assess the relative stability 
of ice polymorphs~\cite{EngelNeeds2015}.

What we need to address is the magnitude of the contribution of these qualitatively 
different components to the  absolute and relative free energies. 
Recent work on water and other hydrogen-bonded systems~\cite{LiMichaelides2011, mark-bern12pnas, mcke+14jcp, RossiMichaelides2015, ceri+16cr}  has demonstrated
that competing NQE in modes parallel and perpendicular to the hydrogen bonds 
partially compensate, reducing their net contribution to free energies. 
In compounds of pharmaceutical interest where polymorphs are almost
isoenergetic (displaying different H-bond networks), NQE can have an important impact.
Anharmonic contributions can have markedly different magnitudes depending
on the system in question -- and for organic molecules containing soft modes,
they are expected to play an important role \cite{RossiBlum2013}. 
As we will see, the magnitude of these contributions 
is strongly connected with the underlying potential energy
surface, be it based on empirical potentials or on
different electronic structure methods.

\begin{figure}[htbp]
\centering
\includegraphics[width=0.45\textwidth]{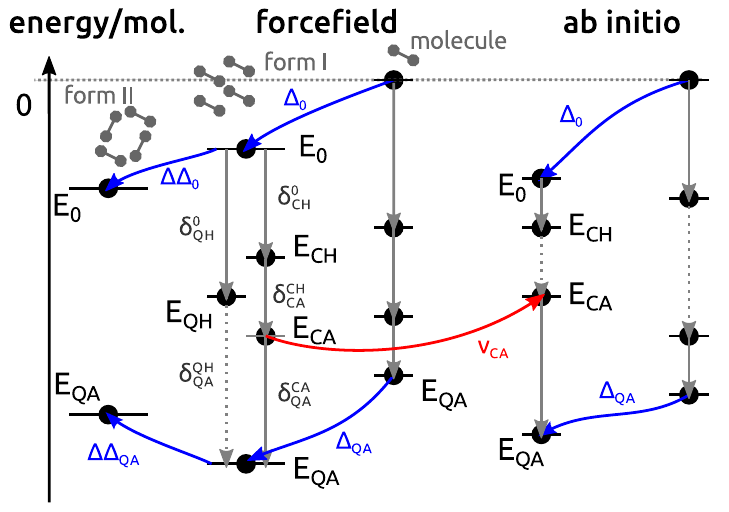}
\caption{A schematic representation of the many components
to the cohesive free energy of molecular crystals. We indicate
with $E_\text{X}$ the (free) energy of a state at a given level of theory:
$E_0$ indicates the bare potential energy, $E_\text{CH}$ the 
classical free-energy computed using a harmonic approximation,
$E_\text{QH}$ the quantum harmonic free-energy, $E_\text{CA}$ the 
classical free-energy using the full anharmonic potential,
$E_\text{QA}$ the quantum anharmonic free-energy. Furthermore, we label
$\delta^\text{X}_\text{Y}$ the energy difference between
levels of theory (e.g. $\delta^\text{CH}_\text{CA}=E_\text{CA}-E_\text{CH}$), 
with $\Delta_\text{X}$ the cohesive energy computed at
a given level of theory (e.g. $\Delta_\text{CA}=E_\text{CA}(\text{fI,FF})-E_\text{CA}(\text{mol,FF})$), 
with $\Delta\Delta_\text{X}$ the relative stability of 
the two forms (e.g. $\Delta\Delta_\text{QA}=\Delta_\text{QA}(\text{fI,FF})-\Delta_\text{QA}(\text{fII,FF})$), 
and finally with $v_\text{X}$ the variation in free-energy
upon changing the potential energy surface. 
}
\label{fig:scheme}
\end{figure}

Our approach to disentangle these contributions relies on 
the combination of multiple thermodynamic transformations,
as depicted schematically in Fig.~\ref{fig:scheme}, where
we also introduce the notation \mc{we will use throughout this work} to refer to the 
many components of the binding free energy.
For harmonic free energies, it is sufficient to evaluate the vibrational frequencies 
for the system at hand and use textbook expressions for a simple harmonic oscillator. 
The difference between the value obtained with quantum and classical oscillators,
$\delta^\text{CH}_\text{QH}$ gives a first idea of the importance
of nuclear quantum fluctuations. Evaluating the anharmonic
contributions with an ab initio potential is instead an
authentic ``tour de force'', that required around two million
CPU hours on a high-performance computing system, despite 
the fact that we deployed an array of acceleration 
techniques, as we will briefly summarize below (see also the SI). 

We computed $\delta^\text{CH}_\text{CA}$ by thermodynamic 
integration from the Debye crystal to the full 
potential~\cite{HabeMano2011}.  Even when using a Debye reference,
the integrand exhibits a near-singularity when approaching the
full potential, probably 
due to the presence, 
of near-zero frequency librations of the methyl groups~\cite{tsap+14mp} -- that instead appear
as finite-frequency vibrations in the harmonic approximation (HA). 
To obtain a converged value for this term we had to (1) use
a ``poor man's self-consistent phonons'' reference, in which
we computed the Hessian using finite differences adapted
to thermal fluctuations in the various directions; 
(2) use a highly non-uniform integration grid;
(3) use a Pad{\'e} interpolation to perform the integral,
which can be motivated by considering the expression for 
the integrand in the case of two (different) harmonic 
reference potentials. Despite these stratagems, it would be
still impractical to compute $E_\text{CA}$ for an \emph{ab initio} (AI)
potential. %
Instead, we first computed $E_\text{CA}(\text{FF})$, and then performed a thermodynamic
integration switching from the force field (FF) to the first-principles
potential, which gives us access to $v_\text{CA}$ (see Fig. \ref{fig:scheme}) 
with a better-behaved integrand, and then computing 
$E_\text{CA}(\text{AI})=E_\text{CA}(\text{FF})+v_\text{CA}$.
The last term one needs to compute is the classical to quantum
change in free energy, $\delta^\text{CA}_\text{QA}$. This can
be obtained by performing a mass thermodynamic integration (MTI), evaluating the quantum kinetic energy of systems
with scaled masses, as obtained from path integral molecular 
dynamics \cite{RossiMichaelides2015, FangMichaelides2016, isotope}. Again, in order to make this calculation feasible
on an \emph{ab initio} potential, we had to
combine all the tricks of the trade: We used (1) an integration
variable that regularizes the integrand~\cite{isotope};
(2) a generalized Langevin equation to reach convergence
of the quantum kinetic energy using only six path integral 
replicas~\cite{ceri-mano12prl}; (3) a multiple time step (MTS)
integrator~\cite{tuck+92jcp}, alternating force 
evaluations with a more converged
basis set and a cheaper, less-converged one. Details of this
scheme and its convergence are discussed 
in the SI. 

In order to perform the steps summarized above, we employed
the i-PI program \cite{ceri+14cpc}, with its recently-developed
MTS implementation~\cite{kapi+16jcp}, 
and its interfaces to LAMMPS \cite{plim95jcp} and CP2K \cite{CP2K,vand+05cpc} 
in order to obtain the FF and DFT energies and forces, respectively.
The FF parameters were generated through the 
SwissParam server~\cite{zoete2011swissparam}, which is based on the Merck molecular force field (MMFF)~\cite{halgren1996merck}.
For the DFT evaluations, we used the PBE \cite{perd+96prl} functional, combined with Goedecker-Teter-Hutter (GTH) pseudopotentials~\cite{goed+96prb} and a TZV2P basis sets \cite{CP2K}. We included the D3 dispersion correction
\cite{grim+10jcp} for vdW interactions. The dynamics and vibrational frequency calculations were performed with only $\Gamma$ point sampling (see SI for a discussion on k-point sampling), and unless otherwise specified, all molecular 
dynamics simulations have been performed at 300K,
using an optimal-sampling Generalized Langevin thermostat\cite{ceri+10jctc}. All free-energies
will be quoted at this temperature. The crystal structure of both forms I and II were 
taken from the CCDC~\cite{AllenCSD2002} and we
kept the lattice parameters fixed at the experimental values\mc{, which were both measured at 300K. 
Optimizing the cell parameters did not change the relative energetics (see SI).}

\begin{table}
\begin{tabular}{c | c c c | c c c}
\hline\hline
 [meV]         &  fI (FF)  &  fII (FF) & $\Delta\Delta$ &  fI (AI)   & fII (AI)  & $\Delta\Delta$  \\
 \hlineˇ
$\Delta_0$             &  -1488   &  -1487 &     -1     &  -1492     &    -1489     &    -3       \\
$\Delta_\text{CH}$     &  -1486   &  -1508 &     22     &  -1500     &    -1487     &   -13             \\
$\Delta_\text{CA}$     &  -1097   &  -1120 &     23     &  -1155     &    -1118     &   -38          \\
$\Delta_\text{QH}$     &  -1451   &  -1474 &     23     &  -1497     &    -1479     &   -18         \\
\bf{$\Delta_\text{QA}$}&\bf{-1060}& \bf{-1082}  &\bf{ 22}  &\bf{-1152}  &\bf{-1107}&\bf{-45}  \\
\hline
 \end{tabular} 
\caption{\label{tab:results}
Binding energies for form I and form II of paracetamol, in meV/molecule, 
computed at different levels of theory, using an empirical force field and 
\textit{ab initio} DFT-PBE+D3 simulations. $\Delta \Delta$ is the relative free-energy difference 
$\Delta(\text{fI})-\Delta(\text{fII})$. 
}
\end{table}

Table~\ref{tab:results} reports the formation energies $\Delta_\text{\text{X}}$ for the two
phases at different levels of theory, as well as the 
difference in stability between form I and form II ($\Delta \Delta$), 
based on an empirical force field (FF) and on {\it ab initio} (AI)
calculations. Perhaps the most striking feature is
that for both FF and AI calculations the two forms
have a very similar binding energy $\Delta_0$,
and that the key to understanding their relative
stability lies in the quantum and finite-temperature
contributions. 

For the finite-temperature contributions, FF simulations consistently 
misrepresent the relative stability of the two forms, making form II
more stable than form I.
Although different contributions change very 
significantly the binding free energy -- the most 
dramatic being the anharmonic classical free energy
term, that destabilizes the crystal by almost
400 meV -- the changes are not reflected on the relative
stability of the two forms, that is constant within
1 meV.
AI data tells a different story. 
Form I is predicted to be the most stable structure,
in agreement with experimental observations. Contrary
to the FF case, the precise value of the 
free-energetic stability depends on a delicate balance
between all the terms, with anharmonicity and 
quantum effects playing a crucial role. In this
case, the classical, harmonic approximation would
suffice to predict the most stable form.
However, anharmonic and nuclear quantum effects contribute by as much
as 30 meV/molecule. 
We can thus infer that the FF simulations fail 
to grasp the differences in the physical effects 
brought by the different molecular stacking and H-bond
pattern of these two forms. 

\begin{figure*}
\centering
\includegraphics[width=1.0\textwidth]{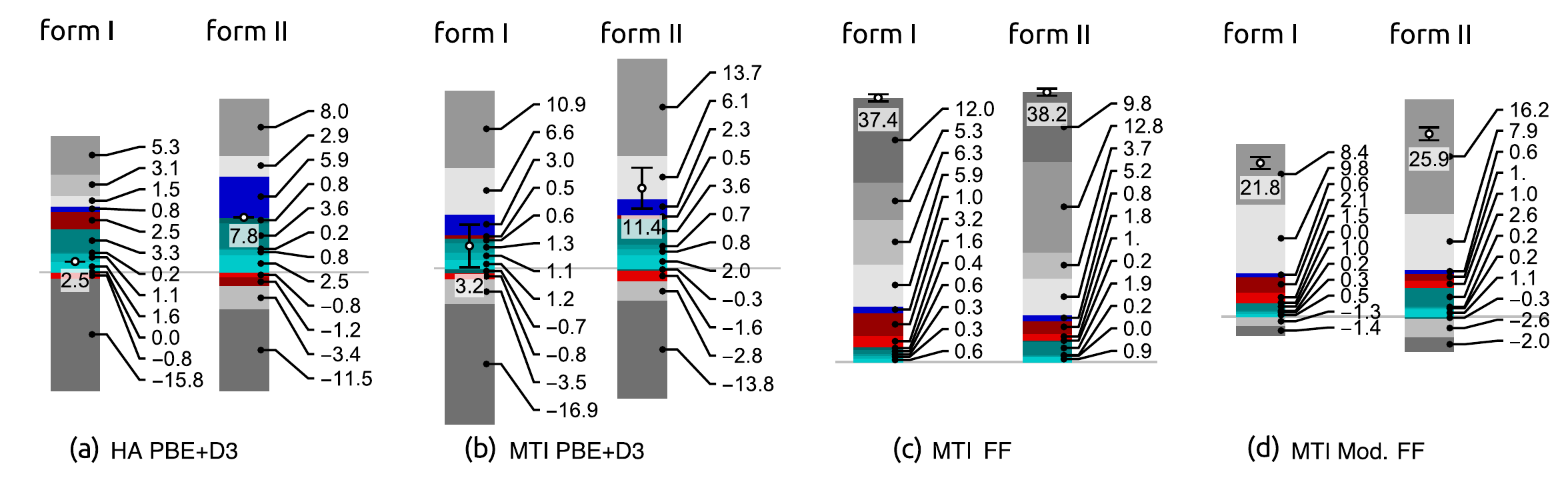}
\caption{Bar plots showing the contribution of each atom (color coded as in Fig. \ref{fig:para}(a)) and the total sum (black circles) of the quantum contributions to the free energies for form I and form II of paracetamol. \mc{All numbers are in meV/molecule.} a) PBE+D3 functional, harmonic approximation $\Delta_\text{QH}-\Delta_\text{CH}$, b) PBE+D3 functional, mass thermodynamic integration $\Delta_\text{QA}-\Delta_\text{CA}$, c) orce field, mass thermodynamic integration $\Delta_\text{QA}-\Delta_\text{CA}$, d) force field with reparametrized anharmonic OH, NH,and CH stretches, mass thermodynamic integration $\Delta_\text{QA}-\Delta_\text{CA}$.}
\label{fig:all-contributions}
\end{figure*}

NQE deserve a more in-depth 
analysis -- both because they are typically neglected,
and because they show strikingly different behavior
in the FF and AI calculations. In the FF, the quantum
contributions to the free energies in the harmonic approximation ($\Delta_\text{QH}-\Delta_\text{CH}$)
amount to 35 and 33 meV for forms I and II respectively. Including
anharmonic effects ($\Delta_\text{QA}-\Delta_\text{CA}$), these quantum contributions are slightly more positive (destabilizing
the crystals), amounting to 37 and 38 meV respectively. In the AI
simulations, instead, the quantum contributions 
are much smaller and differ more significantly between the two forms.
At the harmonic level quantum corrections amounts to only 2 and 8 meV
 for forms I and II, and 3 and 11 meV in the anharmonic case. 
 In order to gain more insight into the role of quantum fluctuations,
we analyzed the contribution of each atom in the paracetamol molecule to the binding quantum
free energy. 
The PBE+D3 results (Figure \ref{fig:all-contributions} (a) and (b)) show that the small
quantum contribution to the binding energy arises from a competition between
quantum effects, as seen in many H-bonded systems~\cite{li+11pnas,ross+15jpcl}.
The H-bonded hydrogen atom gives a significant stabilizing contribution, whereas
most of the heavy atoms, as well as the non-bonding hydrogens, 
destabilize the crystalline phases. 
The stabilizing and destabilizing contributions from different atomic species
sum up to small total values. Even if the details from the atomic 
contributions in the harmonic and anharmonic case are slightly different, 
these numbers are very small and similar, and we can only conclude that 
-- at least in this case -- the harmonic approximation
is sufficient to grasp all the necessary quantum contributions to the free energies. 
It is interesting that form II is more destabilized by
the quantum contributions than form I, and that those contributions make up for around a quarter of the free energy
differences we observe between the two forms.

Performing a similar analysis for the FF simulations 
(Fig. \ref{fig:all-contributions}(c)) 
explains why this inexpensive model differs so much from DFT.
In contrast to the AI data, we find all contributions to 
be positive. The H-bonded atoms also engender a destabilizing effect 
upon binding. Inspired by the observation of the crucial role played by 
an anharmonic parameterization of the OH stretch in the description of 
NQE in water~\cite{mark-bern12pnas}, we 
parametrized fourth order polynomial potentials for the OH, NH, and CH 
stretches based on our PBE+D3 simulations for the isolated molecule (reported in the SI).
We then recalculated the quantum anharmonic contributions with these modified parameters and functional form, which we show in Fig. \ref{fig:all-contributions}(d). 
This simple modification makes the contribution of the OH and NH hydrogens
negative (stabilizing), reducing $\Delta_\text{QA}-\Delta_\text{CA}$ by about 10 meV
for both forms. 
This effect can be explained based on the fact that stabilizing nuclear 
quantum contributions to the cohesive free energy arise due to modes that 
are red-shifted (softened) upon binding. The harmonic functional form of the
FF does not allow the OH and NH stretches to soften sufficiently in 
the crystal forms, while even a simple anharmonic term allows for a 
significant degree of softening.

Although the main focus of this paper is a detailed analysis of free-energy corrections to
the stability of molecular crystals, it is also necessary
to discuss the role of the electronic-structure model. 
We refer the reader to Table II of the SI for the details of this analysis,
where we compare different functionals, k-point meshes, and cell sizes. 
The most important points are that 
\mc{(i) Brillouin Zone sampling has a noticeable 
impact on binding energies, although, contrary
to other examples \cite{NymanGraeme2015}, $\Gamma$-point sampling
suffices for relative energetics.}
(ii) Exact exchange corrections
seem to play a minor role in determining the relative stability of form I and
form II. The PBE0 functional~\cite{pbe0} with the Grimme D3 vdW correction (PBE0+D3) predicts
both forms to be isoenergetic ($\Delta\Delta_0=0$ meV), similar to the PBE+D3 prediction.
(iii) Changing the form of the pairwise vdW correction has a more sizable 
effect. Using the Tkatchenko-Scheffler pairwise vdW correction~\cite{tkat-sche09prl} on top of the
PBE functional (PBE+TS-vdW) stabilizes form II over form I by $\Delta\Delta_0=8$ meV.
(iv) Using a many-body vdW correction~\cite{TkatchenkoScheffler2012,TkatchenkoDistasio2013} can change the energetics significantly, 
but the exact value depends subtly on supercell
size. When using a large supercell, there is 
a small further stabilization of form II, yielding $\Delta\Delta_0=11$ meV (in good agreement with Ref. \cite{DistasioTkatchenko2012}).
(v) Classical and quantum harmonic free energy corrections are relatively
transferable if using pairwise dispersion energy corrections, but are not fully with many body dispersion. 
When considering a double unit cell for form I with the PBE+MBD functional, phonon contributions stabilize it by 3 
(classical) and 10 (quantum) meV. However, if considering a single unit unit cell, form I is {\it destabilized} by 10 (classical) and 2 (quantum) meV. This observation is consistent with long 
range fluctuations described by this method that were reported in the literature \cite{ReillyTkatchenko2014, Ambrosetti1171}

\mc{The detailed analysis we 
performed for paracetamol}
underscores the grand challenge 
that is faced by efforts to predict the 
most stable polymorph from first 
principles \cite{Bardwell:2011}. 
Subtle, hard-to-compute terms such as the anharmonic free energy or NQE contribute
by similar amounts to the overall energy balance, 
and any estimate that does not include the full
package risks obtaining the right result for the 
wrong reason, or just a plain wrong result. 
To complicate things further, we observe a strong
interplay between these free-energy terms and the underlying
potential energy surface. A harmonic FF would
incorrectly predict that the classical harmonic terms 
are enough to correct the baseline relative energy 
$\Delta\Delta_0$, whereas AI energetics show that 
anharmonic contributions are of 
paramount importance, and that 
quantum effects involve an 
almost complete cancellation between
competing terms. The details of the 
AI calculation are also
important: the kind of pairwise dispersion 
interactions and the use of many body dispersion corrections can change
the relative stability of different polymorphs by tens of meV. 
\mc{ In this case, it appears that
one can forgo expensive 
exact exchange calculations, but results
on different materials suggest that this observation
is not universal~\cite{MaromTkatchenko2013}.}
At the harmonic level, 
pairwise corrections predict form I to be more stable,
in agreement with experiments \cite{Espeau2005524}.
MBD, which includes more physics and is generally
more accurate, would apparently need the full anharmonic
treatment in order to grasp this energetic balance --
but the computation of these terms would
be prohibitively expensive at this point. 

\mc{The fact that anharmonic and quantum effects 
can be important also for a relatively
simple molecule hints at the challenges that
will be faced as CSP ventures into molecules
with greater conformational flexibility.
While great progress has been made towards
reliable electronic-structure calculation of 
binding energies \cite{NymanGraeme2015, KronikTkatchenko2014, BrandenburgGrimme2014}, 
the same attention should 
now be given to the evaluation of anharmonic 
and quantum free energy -- to avoid painting 
an incomplete, possibly misleading picture.}
\mc{Future work shall investigate systematically
these effects in different classes of molecular
crystals, and benchmark different approximations
to compute them inexpensively. Our study 
already provides 
hints at how to achieve predictive accuracy, 
without paying an unreasonable price.}
Computing anharmonic free energies
through an indirect route that exploits
integration from a FF reference
greatly reduces the effort. 
It appears that even in the presence of
competing quantum effects, 
$\delta^\text{CH}_\text{QH}$ 
gives a good estimate of the 
anharmonic quantum correction $\delta^\text{CA}_\text{QA}$, 
provided that the underlying 
potential energy surface 
can capture the environment-dependent
softening of H-bonds. In that respect,
augmenting empirical force fields to
include anharmonic corrections to the
bond energies could increase their 
reliability in predicting the impact of 
NQE. 
The path to an \emph{ab initio} prediction 
of the stability of \mc{complex} molecular
crystals is ripe with challenges, but
the stakes are high. With the combined
progress in searching complex structural
landscapes, predicting accurately
vdW interactions, and efficiently
estimating classical and quantum
free-energy contributions, this 
goal is getting near.

\begin{acknowledgments}
We would like to thank V. Kapil for assistance
in the implementation of normal-mode
analysis in i-PI. 
We acknowledge generous allocation of CPU time by 
CSCS under the project id s618.
This research was (partly) supported by the NCCR
MARVEL, funded by the Swiss National Science 
Foundation, the SNF project 
200021-159896 and the MPG-EPFL Center for Molecular 
Nanoscience. 
MR acknowledges funding by the Max Planck Society.
\end{acknowledgments}
\bibliographystyle{apsrev4-1}

\end{document}